\documentclass[lettersize,journal]{IEEEtran}
\usepackage{amsmath,amsfonts}
\usepackage{algorithmic}
\usepackage{algorithm}
\usepackage{array}
\usepackage{amsmath}
\usepackage[caption=false,font=normalsize,labelfont=sf,textfont=sf]{subfig}
\usepackage{textcomp}
\usepackage{stfloats}
\usepackage{url}
\usepackage{verbatim}
\usepackage{graphicx}
\usepackage{cite}
\usepackage{mathtools}
\usepackage{color}
\newcommand{\argmin}{\mathop{\rm arg~min}\limits}

\begin{document}
	
	\title{\textcolor{black}{Intuitive Directional Sense Presentation to the Torso Using McKibben-Based Surface Haptic Sensation in Immersive Space}}
	
	\author{Kenta Yokoe,~\IEEEmembership{Graduate Student Member,~IEEE,} Tadayoshi Aoyama,~\IEEEmembership{Member,~IEEE,} \\Yuki Funabora,~\IEEEmembership{Member,~IEEE,}  Masaru Takeuchi,~\IEEEmembership{Member,~IEEE} and Yasuhisa Hasegawa,~\IEEEmembership{Member,~IEEE}
		\thanks{This work was supported by JST [Moonshot R\&D] Grant Number [JPMJMS2214-08], JSPS KAKENHI Grant Number JP22H03630 and JP23KJ1116, and Program for Promoting the Enhancement of Research Universities.}
		\thanks{The protocol of this study was approved by the Ethics Committee of the Graduate School of Engineering, Nagoya University (Approved Number: 22-9).}
		\thanks{K. Yokoe, M. Takeuchi, and Y. Hasegawa are with the Department of Micro-Nano Mechanical Science and Engineering, Nagoya University, Nagoya, Aichi, 464-8603, Japan.
			{e-mail: yokoe@nagoya-u.jp, tadayoshi.aoyama@mae.nagoya-u.ac.jp,  masaru.takeuchi@mae.nagoya-u.ac.jp, hasegawa@mein.nagoya-u.ac.jp}}%
		\thanks{K. Yokoe is a Research Fellow of Japan Society for the Promotion of Science. }%
		\thanks{T. Aoyama is with the Department of Mechanical Systems Engineering, Nagoya University, Furo-cho, Chikusa-ku , Nagoya, Aichi 464-8603, Japan, and the Center for One Medicine Innovative Translational Research, Gifu University, Yanagido 1-1, Gifu, Gifu 501-1193, Japan.
			{e-mail: tadayoshi.aoyama@mae.nagoya-u.ac.jp}}
		\thanks{Y. Funabora is with the Department of Information and Communication Engineering, Nagoya University, Nagoya, Aichi, 464-8603, Japan.
			{e-mail: funabora@nagoya-u.jp}}%
		\thanks{Manuscript received 6 December 2023; revised 1 August 2024 and 31 October 2024; accepted 15 December 2024.}
		\thanks{This is the accepted version of an article published in IEEE Transactions on Haptics, vol.~18, no.~1, pp.~244--254, 2025, DOI: 10.1109/TOH.2024.3522897. \copyright~2024 The Authors. Open Access under CC BY-NC-ND 4.0 (https://creativecommons.org/licenses/by-nc-nd/4.0/).}
	}
	
	\markboth{IEEE TRANSACTIONS ON HAPTICS}%
	{Yokoe \MakeLowercase{\textit{et al.}}: Intuitive Directional Sense Presentation to the Torso Using McKibben-based Surface Haptic Sensation in Immersive Space. }
	
	\IEEEpubid{0000--0000/00\$00.00~\copyright~2024 IEEE}
	
	\maketitle
	
	\begin{abstract}
		In recent years, systems that utilize immersive space have been developed in various fields. 
		\textcolor{black}{
			Immersive spaces often contain considerable amounts of visual information; therefore, users often fail to obtain their desired information. 
			Therefore, various methods have been developed to guide users toward haptic sensations.
			However, many of these methods have limitations in terms of the intuitive perception of haptic sensation and require practice for familiarization with haptic sensation.
		}
		Fabric actuators are wearable haptic devices that combine fabric and McKibben artificial muscles to provide wearers with surface haptic sensation. 
		These sensations can be provided to a wide area of the body \textcolor{black}{with intuitive perception}, instead of only to a part of the body.
		This paper presents a novel air pressure adjustment method for whole-body motion guidance using surface haptic sensations provided by a wearable fabric actuator. The proposed system can provide users with a directional sense without visual information in an immersive space. 
		The effectiveness of the proposed system was evaluated through subject experiments and statistical data analysis. 
		Finally, a directional sense presentation was conducted for users performing micromanipulations in a mixed-reality space to demonstrate the applicability of the proposed system for teleoperation.
	\end{abstract}
	
	\begin{IEEEkeywords}
		\textcolor{black}{
			Wearable haptics, McKibben artificial muscle, Kinesthetic sensation, Directional guidance, Immersive environments
		}
	\end{IEEEkeywords}
	
	\section{Introduction}
	\textcolor{black}{
		Recently, the demand for systems based on immersive spaces has increased \cite{lipton2018Baxter}. 
		In the medical field, immersive spaces have been utilized in applications such as medical education  \cite{chen2017Recent}, surgical planning \cite{verhey2020Virtual}, surgical guidance \cite{gregory2018Surgery},  and assisted reproductive technology \cite{aoyama2020HeadMounted,yokoe2022immersive}.  
	}
	Users in immersive spaces can obtain rich visual information in all directions. 
	\textcolor{black}{
		However, excessive visual information may cause users to overlook their desired information.
	}
	This issue is especially prevalent in unknown spaces, such as micro space, where users often look in unintended directions owing to insufficient information regarding the objects in the space.
	\begin{figure}[!t]
		\centering
		\includegraphics[keepaspectratio=true,width=\linewidth]{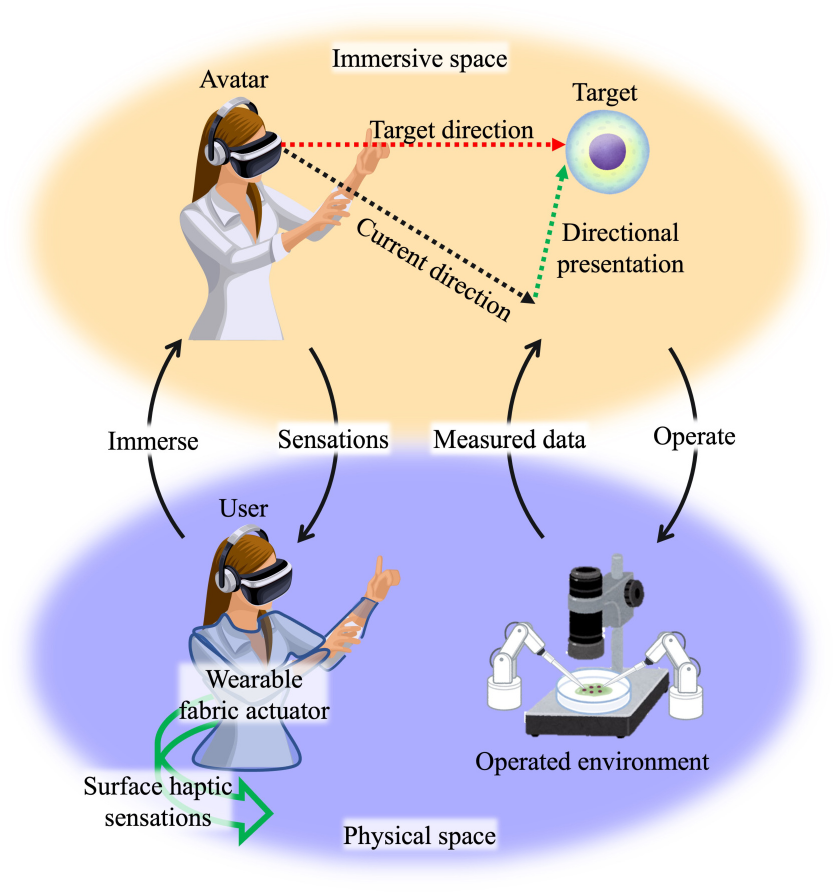}
		\vskip -3mm
		\caption{Conceptual schematic of the directional sense presentation system. \textcolor{black}{The top and bottom are the immersive and physical spaces, respectively. The user is immersed in the immersive space consisting of the operating environment and performs operations.}}
		\label{fig:concept}
	\end{figure}
	
	\IEEEpubidadjcol
	\textcolor{black}{
		To address this problem, guidance methods based on visual information have been developed for immersive spaces \cite{lee2020Visual,grogorick2017Subtle}. 
		These methods guide user movements by adding or changing visual information. 
		Users of these methods must closely observe visual information. 
		However, as close observation of visual guidance can interfere with the user's work in an immersive space, guidance methods that add or change visual information are undesirable.
		Therefore, a directional guidance method that does not require visual information is required. 		
	}
	
		This study aimed to provide directional guidance to users without visual information. 
	We focused on haptic sensation instead of visual information as a means of directional guidance.
	\textcolor{black}{
		When humans move their gaze, their head and torso movements are accordingly synchronized \cite{sidenmark2020Eye}.
		The torso movement is particularly important during large gaze movements.
		In addition, when guiding the user's gaze,  the object that the user wants to see must be placed in the central vision field, which is 30 deg in radius from the center of the viewpoint~\cite{sampson2012Visual,bhise2011Ergonomics}.
		Therefore, we propose a method to indirectly guide a user's gaze by changing the posture of the torso in an immersive space using haptic sensation to place the object that the user wants to see in the user's central field of view.
		We selected McKibben-based surface haptic sensations provided by wearable fabric actuators to change the torso posture of the user. The fabric actuators are composed of clothing and McKibben-type artificial muscles.  
		The haptic sensation is provided to the user by deforming clothing with McKibben-type artificial muscles attached to fabric actuators \cite{peng2023FunabotSuit}. 
		This haptic sensation intuitively provides a user's torso with four kinesthetic perceptions: forward bending, backward bending, left twist, and right twist. 
	}
	
	\textcolor{black}{Fig.~\ref{fig:concept} shows a schematic of the proposed system. }
	The user is immersed in an immersive space created by measured data from the operated environment.
	\textcolor{black}{
		The proposed system provides surface haptic sensations to users based on their current and target gaze directions to guide their gaze in an immersive space.
		In this study, we propose a method for indirectly guiding a user's gaze by providing surface haptic sensations that induce torso movement. 
		The user alters the posture of the torso according to the surface haptic sensation, and adjusts the head and eye positions according to the torso posture.
		The proposed system implements a novel method for adjusting air pressure based on the user's gaze direction and the target object that the user wishes to see. 
		We conducted a subject experiment to confirm whether the proposed system could display the target object that the user wanted to see in the central field of view.
		In addition, a subject experiment was conducted to evaluate the perceived speed of the surface haptic sensation.
		Finally, the proposed system was demonstrated in a micromanipulation scenario conducted in an actual immersive space.
	}

	\section{Related Works}
	Several devices have been developed to enable the presentation of haptic sensations instead of visual information in immersive spaces \cite{wang2020Multimodal}. 
	\textcolor{black}{
		Some haptic sensation methods in immersive spaces use stationary devices such as friction, hardness, and roughness \cite{culbertson2017Importance,yang2014Enhancing}.
	}
	However, these devices have limitations such as a lack of space for presentation because 
	\textcolor{black}{
		they are fixed at one location. 
	} 
	To address these issues, wearable and lightweight haptic sensation devices have been developed recently \cite{pacchierotti2017Wearable}.
	\textcolor{black}{
		Many of these wearable devices focus on the hands and fingers. 
		These devices improve the reality of the immersive space by providing the sensations of softness \cite{blake2009Haptic}, pressure \cite{lim2021Development}, and touch \cite{lee2019Wearable,wu2017virtual,edwards2019Haptic} via haptic sensation to the hand.
		However, methods that focus on the hands and fingers provide limited haptic information because of the limited areas of these parts.
	} 
	\textcolor{black}{
		To provide more haptic information to the user, approaches that provide haptic sensations to various parts of the body are also available. 
		For example, haptic sensations can be provided to the arms \cite{schonauer2012Multimodal}, feet \cite{meier2015Exploring}, neck \cite{yamazaki2019Neck,tanaka2022Electricalb}, head \cite{yu2023DrivingVibe,ke2023TurnAhead}, face \cite{nakamura2022Evaluation}, torso \cite{tsai2022ImpactVest}, and at various body locations \cite{al-sada2020HapticSnakes}.
	}
	
	\textcolor{black}{
		Haptic sensations to various parts of the body enable tasks such as navigation and guidance of movements in immersive spaces, which are difficult with haptic presentation using only the hands and fingers \cite{meier2015Exploring,tanaka2022Electricalb,nakamura2022Evaluation}. 
		These methods use haptic sensations to the face and neck to provide gaze guidance.
		However, these methods do not consider the movement of the torso, which significantly influences large gaze movements \cite{sidenmark2020Eye}, as described in the Introduction.
		Other similar approaches provide navigation and guidance through haptic sensation to the user's torso \cite{li2013Design,monica2023Improving,ertan1998wearable}.
		These methods enable navigation by providing vibrotactile sensations to the torso.
		However, these methods require prior instructions for the user on the vibrotactile sensations that indicate a specific type of navigation, making these systems less intuitive for first-time users.
		Additionally, these methods are specialized for two-dimensional (2D) navigation on horizontal maps, making their application to  three-dimensional (3D) gaze guidance difficult.
	}
	
	\textcolor{black}{Unlike vibrotactile sensations, McKibben-based}
	surface haptic sensations provided by fabric actuators can present haptic sensations to a wide area of the body
	\textcolor{black}{
		using intuitive haptic perception \cite{funabora2017Prototype,funabora2018Flexible,peng2023FunabotSuit}. 
		Such actuators use the deformation of clothing by McKibben's artificial muscles as a haptic sensation, allowing the user to intuitively interpret the information that the haptic sensation intends to provide.
		Our approach uses this haptic sensation to enable 3D gaze guidance with intuitive haptic perception.
	}

	
	\section{Physical Directional Sense Presentation System Using a Wearable Fabric Actuator}
	\label{sec:system}
	\begin{figure}[!t]
		\centering
		\includegraphics[keepaspectratio=true,width=\linewidth]{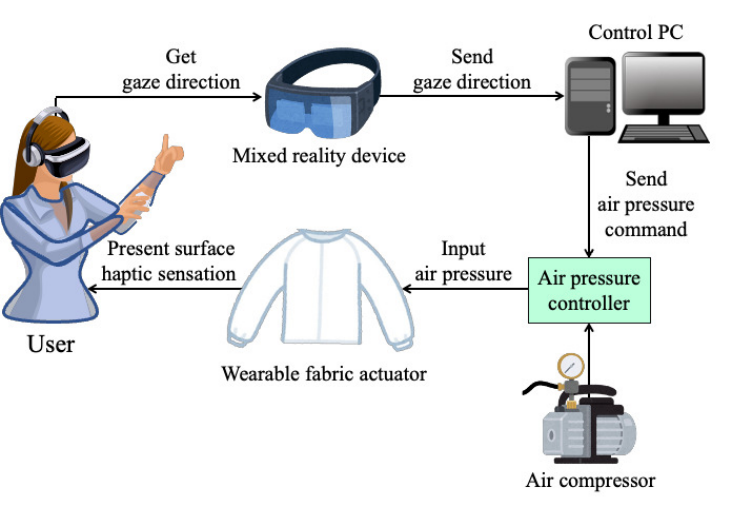}
		\vskip -3mm
		\caption{Configuration of the directional sense presentation system. \textcolor{black}{The user wears a mixed reality device with gaze tracking and our wearable fabric actuator. The fabric actuator provides surface haptic sensations based on the user's gaze received by the mixed reality device.}}
		\label{fig:fabric_config}
	\end{figure}
	\begin{figure}[!t]
		\centering
		\includegraphics[keepaspectratio=true,width=.9\linewidth]{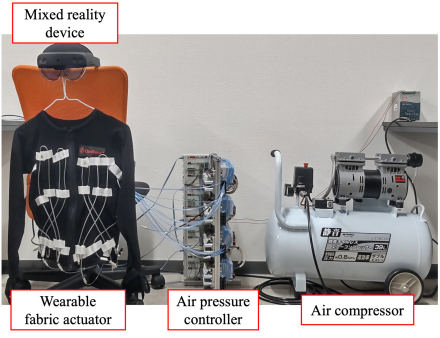}
		\vskip -3mm
		\caption{Overview of the directional sense presentation system. }
		\label{fig:fabric_overview}
	\end{figure}
	Figs.~\ref{fig:fabric_config} and \ref{fig:fabric_overview} show the configuration and overview of the physical directional sense presentation system using a wearable fabric actuator, respectively. 
	The user wears a fabric actuator and mixed-reality device with a gaze-tracking function. Information on the gaze direction from the mixed-reality device is sent to the PC, which calculates the air pressure command values necessary to guide the wearer's gaze and sends them to the air pressure controller.
	The proposed system consists of a wearable fabric actuator, a computer (Windows 10 Pro, 64-bit, CPU Intel (R) Core (TM) i7-11800H, 2.30 GHz, 16 GB RAM, GPU NVIDIA GeForce RTX 3060 Laptop), an air compressor (ACP-39SLA, TAKGI Corporation), an air pressure controller, and a mixed-reality device (Hololens2, Microsoft). The air pressure controller includes programmable logic controllers (007001001100, Industrial shields) and electric regulators (CRCB-0135W/0136W, KOGANEI Corporation).
	
	\begin{figure}[!t]
		\centering
		\includegraphics[keepaspectratio=true,width=\linewidth]{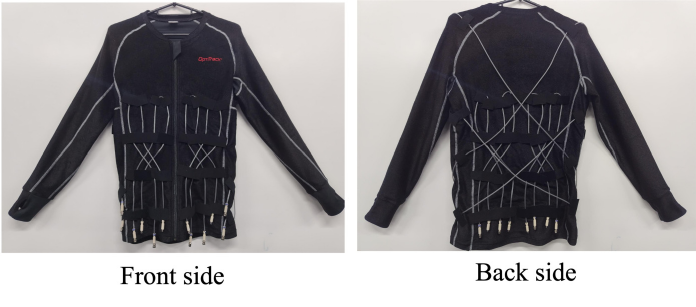}
		\vskip -3mm
		\caption{\textcolor{black}{
				Overview of wearable fabric actuator. The left and right images show the front and back sides, respectively.}}
		\label{fig:suit}
	\end{figure}
	\begin{figure}[!t]
		\centering
		\includegraphics[keepaspectratio=true,width=.9\linewidth]{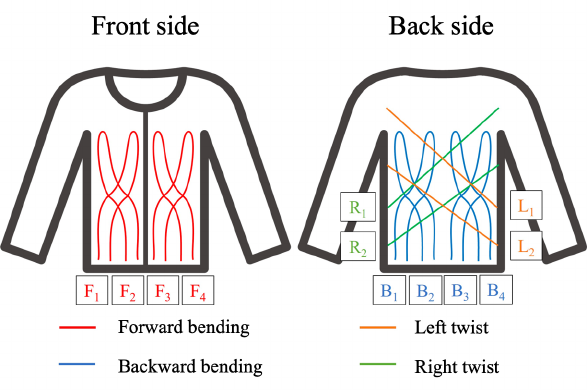}
		\vskip -3mm
		\caption{Placement of twelve McKibben artificial muscles. The left figure shows the front side and the right figure shows the back side. Red, blue, green, and orange lines indicate McKibben artificial muscles. 
		}
		\label{fig:muscle}
	\end{figure}
	Fig.~\ref{fig:suit} presents an overview of the wearable fabric actuator. It consists of 12 McKibben artificial muscles placed on the clothing, which are driven by the air pressure. 
	The arrangement of McKibben artificial muscles in the wearable fabric actuator is illustrated in Fig.~\ref{fig:muscle}. 
	The artificial muscles contracted by the application of air pressure deform the wearable fabric actuator, providing surface haptic sensation to the wearer. The intensity of surface haptic sensations and magnitudes of the applied air pressure and are positively correlated in the wearable fabric actuator.
	By changing the combination of artificial muscles to which air pressure is applied, the wearable fabric actuator can provide the wearer's torso with four sensations, i.e., forward bending, backward bending, left twisting, and right twisting.
	The application of air pressure to the artificial muscles, indicated by red line (F$_1$, F$_2$, F$_3$, F$_4$), blue line (B$_1$, B$_2$, B$_3$, B$_4$), orange line (L$_1$, L$_2$), and green line (L$_1$, L$_2$) in Fig.~\ref{fig:muscle}, produced forward bending, backward bending, left twist, and right twist, respectively.  
	A detailed description of the wearable fabric actuator, including an evaluation of its basic performance, is presented in \cite{peng2023FunabotSuit}.
	
	\section{Algorithm of the Physical Directional Sense Presentation }
	\label{sub:fitting}
	\subsection{Derivation of Equation between Air Pressure and Gaze Angle}
	\textcolor{black}{
		The proposed system divides the gaze angles into two angles on the horizontal (left and right) and vertical (up and down) planes. When the proposed system guides the user's gaze to the left, the fabric actuator provides a surface haptic sensation that induces a left twist of the torso. When the proposed system guides the user's gaze to the right, the fabric actuator provides a surface haptic sensation that induces a right twist of the torso. When the proposed system guides the user's gaze upward, the fabric actuator provides a surface haptic sensation that induces backward bending of the torso. Similarly, when the proposed system guides the user's gaze downward, the fabric actuator provides a surface haptic sensation that induces forward bending of the torso. 
	}
	To achieve air pressure adjustment in surface haptic sensation based on the user gaze angle, it is necessary to determine an equation that relates the air pressure applied to the fabric actuator to the gaze angle.
	This equation must establish a one-to-one correspondence between air pressure and gaze angle, as air pressure values for a target gaze angle cannot be determined without such a correspondence.
	Moreover, the equation requires that the air pressure increases continuously as the gaze angle increases since the surface haptic sensation increases as the air pressure increases.
	Based on these two conditions and Stevens' power law \cite{brintzenhofe2023Investigating}, the relationship between air pressure and gaze angle can be given by: 
	\begin{align}\label{eq:ap}
		P_k = f_k(\theta_k) = a_{k} \theta_k^{b_{k}} + c_{k} \hskip 2 mm (k=1, \cdots, 4), 
	\end{align}
	where $a_k$, $b_k$, and $c_k$ are the coefficients. 
		$k$= 1, 2, 3, and 4 denote left twist, right twist, forward bend, and backward bend, respectively. $P_{k}$ $(k=1, \cdots, 4)$ represents the air pressure applied on the fabric actuator. $\theta_{k}$ $(k=1, \cdots, 4)$ represent the gaze angles in the left, right, down, and up directions.  
	
\begin{figure}[!t]
		\centering
		\includegraphics[keepaspectratio=true,width=\linewidth]{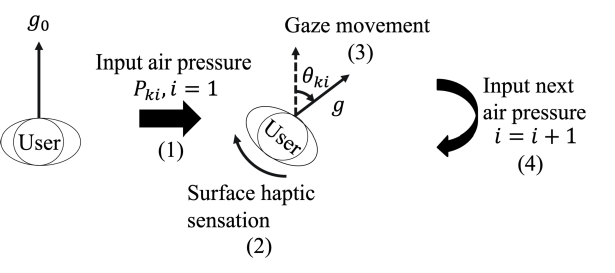}
		\vskip -3mm
		\caption{\textcolor{black}{
				Fitting procedure view from the top. 
				(1): In the initial state where the user is front facing, input the first air pressure into the wearable fabric actuator. 
				(2): The wearable fabric actuator provides surface haptic sensation. 
				(3): The user moves the gaze direction. Obtain the angle between the initial gaze vector $g_0$ and gaze vector $g$. 
				(4): Input the next air pressure.
				Repeat steps (2) to (4) to obtain the combination of $P_{kn}$ $(n=1, \cdots, 10)$ and $\theta_{kn}$ $(n=1, \cdots, 10)$; then, perform the least-squares method.}}
		\label{fig:fittingimage}
		\vskip -2mm
	\end{figure}

	\begin{algorithm}[!t]
		\caption{Algorithm of approximate equation fitting.}
		\label{alg:fitting}
		\begin{algorithmic}[1]
			\renewcommand{\algorithmicrequire}{\textbf{Input:}}
			\REQUIRE  $k=1$ to $4$, $n=1$ to $10$, $0 \leq P_{kn} $, $P_{k10} \leq $~$P_{MAX}$, $P_{kn} < P_{kn+1}$
			\FOR  {$i=1$ to $4$}
			\FOR {$j=1$ to $10$}
			\STATE Input air pressure commands of size $P_{ij}$ into the artificial muscles.
			\STATE Obtain the gaze angle.
			\STATE $\theta_{ij} \gets new value$
			\ENDFOR
			\STATE Find an approximate equation for $P_{in}$ $(n=1, \cdots, 10)$ and $\theta_{in}$ $(n=1, \cdots, 10)$ using the least squares method.
			\ENDFOR
		\end{algorithmic}
	\end{algorithm}
	Algorithm~\ref{alg:fitting} is used to determine coefficients $a_k$, $b_k$, and $c_k$ in Equation \eqref{eq:ap}. 
	\textcolor{black}{
		The algorithm employs a systematic data collection and fitting procedure. In this process, $P_{MAX}$ denotes the maximum air pressure, set at 375 kPa in the proposed system. 
		As shown in Equation~\eqref{eq:ap}, the subscript $k$ denotes the torso movements, while the subscript $n$ represents the sequence number of measurements $(n=1, \cdots, 10)$. 
		For each movement $k$, the algorithm collects a sequence of pressure--angle pairs $(P_{kn}, \theta_{kn})$, gradually increasing the applied pressure and recording the corresponding gaze responses.
		These data points are then used in the Levenberg--Marquardt method to determine the optimal coefficients $a_k$, $b_k$, and $c_k$ for each of the four movements: left twist, right twist, forward bend, and backward bend.}
	Fig. \ref{fig:fittingimage} illustrates the fitting procedure from the top. 
	First, the initial gaze vector of the wearer is recorded before fitting.
	Next, air pressure $P_{k1}$ is applied to the artificial muscle to guide it in the direction of fitting ((1) in Fig.~\ref{fig:fittingimage}). 
	Surface haptic sensations corresponding to the air pressure are then applied to the wearer's torso ((2) in Fig.~\ref{fig:fittingimage}). 
	The proposed system records angle $\theta_{ki}$ between the wearer's gaze vector $g$ and initial gaze vector $g_0$, which changes in response to surface haptic sensation from the wearable fabric actuator ((3) in Fig.~\ref{fig:fittingimage}). 
	After recording angle $\theta_{ki}$, the following magnitude of air pressure is applied to the artificial muscles ((4) in Fig.~\ref{fig:fittingimage}). 
	(2) to (4) in Fig.~\ref{fig:fittingimage} are repeated 10 times to obtain the air pressure $P_{ki}~(i=1, \cdots, 10)$ applied to the artificial muscles and the corresponding gaze angle $\theta_{ki}~(i=1, \cdots, 10)$. 
	\textcolor{black}{
		In this fitting procedure, participants are instructed to adjust their torso posture and gaze direction following the haptic sensations and their torso posture, respectively.
	}
	Let $H_k$, $x_k$, and $y_k$ be defined as follows: 
	\begin{align}\label{eq:let}
		H_k = 
		\left[
		\begin{array}{c}
			a_{k} \theta_{k1}^{b_{k}} + c_{k} \\
			a_{k} \theta_{k2}^{b_{k}} + c_{k} \\
			\vdots \\
			a_{k} \theta_{k10}^{b_{k}} + c_{k} 
		\end{array}	
		\right], 
		x_k =
		\left[
		\begin{array}{c}
			a_k  \\
			b_k  \\
			c_k 
		\end{array}	
		\right], 
		y_k =
		\left[
		\begin{array}{c}
			P_{k1}  \\
			P_{k2} \\
			\vdots \\
			P_{k10}
		\end{array}	
		\right] .
	\end{align}
	The coefficients ($a_k$, $b_k$, and $c_k$) can be estimated using the Levenberg-Marquardt method, as follows: 
	\begin{align}\label{eq:lm}
		\hat{x_k} = \argmin_{x_k} \left( y_k - H_k \right)^2.
	\end{align}
	The proposed algorithm performed this fitting for four guiding directions (forward, backward, left, and right) to obtain the correlation between the air pressure and gaze angle in each direction.
	
	\subsection{Air Pressure Adjustment Method for Surface Haptic Sensations based on Gaze Angle}
	\label{sub:feedback}
\begin{figure}[!t]
		\centering
		\includegraphics[keepaspectratio=true,width=\linewidth]{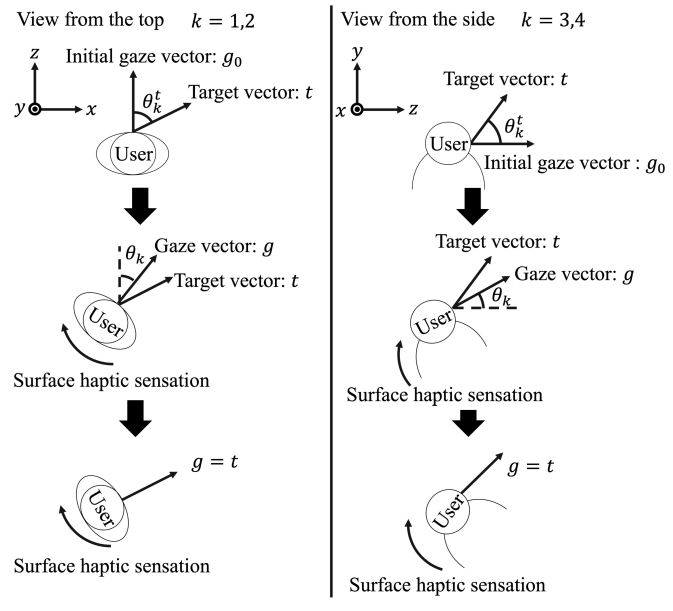}
		\vskip -3mm
		\caption{\textcolor{black}{
				(Left) Procedure for directional sense presentation using the proposed system (top view). (Right) Procedure for directional sense presentation using the proposed system (side view). The left--right directional sense presentation is based on $\theta_1^t$, $\theta_2^t$, $\theta_1$, and $\theta_2$. The up--down directional sense presentation is based on $\theta_3^t$, $\theta_4^t$, $\theta_3$, and $\theta_4$.}
		}
		\label{fig:gazeguidanceimage}
	\end{figure}

	Fig. \ref{fig:gazeguidanceimage} depicts the procedure of directional sense presentation using the proposed system. 
	For left-right guidance, the target angles $\theta_1^t$ and $\theta_2^t$ are calculated from the projection of the gaze vector and target position vector onto the $xz$ plane. The wearable fabric actuator provides surface haptic sensation using the current gaze angles $\theta_1$ or $\theta_2$. 
	For upward and downward guidance, the target angles $\theta_3^t$ and $\theta_4^t$ are obtained from the projection of vectors onto the $yz$ plane. The wearable fabric actuator provides surface haptic sensation using the current gaze angles $\theta_3$ or $\theta_4$. 
	
	The proposed system calculates the target gaze angle to be moved from the gaze and position vectors of the target object relative to the user. 
	The system then calculates the air pressure value as the input from the approximate equation and calculated angle. 
	The current air pressure is fed back based on the user's current gaze angle to reduce disturbances and errors. 
	The target air pressures for the artificial muscles in the four directions ($P_k^t$) are derived as follows: 
	\begin{align}\label{eq:feedback}
		P_k^t  = \left\{
		\begin{array}{ll}
			P_k^t+ G  (f_k(\theta_k^t) - f_k(\theta_k)), \\& \hskip -15mm \mbox{if  ($\theta_k^t > 0$ $\land$ $\frac{d}{dt}(\theta_k^t-\theta_k) > 0$)} \\
			P_k^t, & \hskip -15mm \mbox{if  ($\theta_k^t > 0$ $\land$  $\frac{d}{dt}(\theta_k^t-\theta_k) < 0$)}\\
			0, &\hskip -15mm \mbox{if  $\theta_k^t \le 0$}
		\end{array}
		\right. ,
	\end{align}
	where $\theta_k^t$ is the target gaze angle and $G$ is a constant value.
	The output of surface haptic sensation changes only when the user's gaze moves away from the target gaze direction, and not when it approaches the target gaze direction. 
	\textcolor{black}{
		We believe that when the clothing deformation decreases, the user may misconstrue the gaze angle at which the deformation decreases as the target gaze. Therefore, we adopted an air pressure control algorithm that provides feedback only when the error increases. 	
	}
	
	\section{Experiment to Evaluate Directional Sense Presentation Accuracy }
	\label{sec:accuracy}
	\subsection{Experimental Setup}
\begin{figure}[!t]
		\centering
		\includegraphics[keepaspectratio=true,width=.75\linewidth]{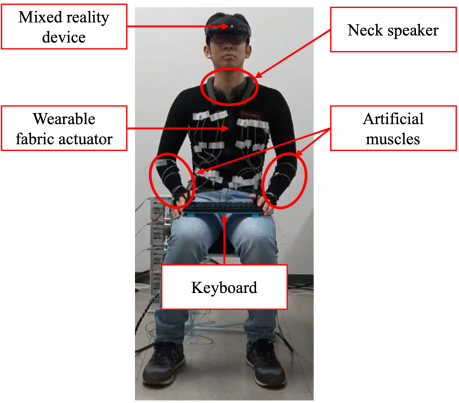}
		\caption{Overview of a subject in the experiment. \textcolor{black}{The subject wears a mixed reality device, proposed wearable fabric actuator, and neck speaker. The subject also has three artificial muscles wrapped around each arm.}}
		\label{fig:mr_subject}
	\end{figure}

	The first subject experiment was conducted to evaluate the accuracy of directional presentation.
	Fig.~\ref{fig:mr_subject} shows the appearance of a subject during the experiment, who wore a mixed-reality device, wearable fabric actuator, neck speaker, and six McKibben-type artificial muscles wrapped around each arm. The subjects sat on a stool and placed a keyboard on their lap. 
	
	Before beginning the experiment, the subjects performed the fitting procedure described in Section \ref{sub:fitting}.
	At the beginning of the experiment, the subject faced forward, and the target gaze angle was randomly determined. It should be noted that the target gaze angle was within 30 deg of the vertical angle and 60 deg of the horizontal angle.
	The subjects changed their gaze angle according to gaze guidance and pressed the keyboard when they thought that the current gaze angle matched the target gaze angle. 
	The accuracy of gaze guidance was evaluated by measuring the error between the target gaze angle and the subjects' gaze angle based on their keyboard input. 
	\begin{table}[!t]
		\caption{Experimental Conditions.}
		\label{table:mr_condition}
		\centering
		\begin{tabular}{c|cc}
			Condition& Guidance method & Air pressure adjustment \\\hline
			1   & Surface haptic sensation & With \\
			2   & Surface haptic sensation & Without \\
			3   & Simple haptic sensation & With \\
			4   & Synthesized speech & Without \\ 
		\end{tabular}
	\end{table}
	\begin{figure}[!t]
		\centering
		\includegraphics[keepaspectratio=true,width=.6\linewidth]{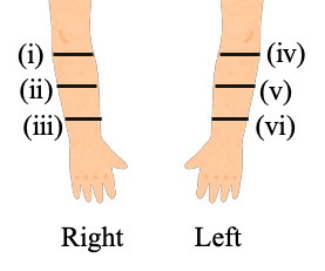}
		\vskip -3mm
		\caption{Arrangement diagram of artificial muscles attached to the arms. The black line indicates the McKibben artificial muscle. These artificial muscles are used to present simple haptic sensations during the experiment.}
		\label{fig:arm}
	\end{figure}
	\begin{table}[!t]
		\centering
		\caption{\textcolor{black}{Relationship between the Gaze Guidance Direction and the Contracted Artificial Muscle in Condition 3.}}
		\label{tab:simple}
		\begin{tabular}{c|c}
			Guiding direction & Contracted artificial muscle in Fig.~\ref{fig:arm}\\ \hline
			Upward    & (i) and (iv)   \\
			Downward  & (iii) and (vi) \\
			Leftward  & (v)            \\
			Rightward & (ii)          
		\end{tabular}
	\end{table}
	The subjects experimented with four conditions using different gaze guidance methods.
	Table~\ref{table:mr_condition} lists the gaze guidance methods used in each experimental condition.
	In Conditions 1 and 2, the surface haptic sensation was used for gaze guidance; the air pressure adjustment method described in Section \ref{sub:feedback} was implemented in Condition 1, but not in Condition 2.
	In Condition 3, directional guidance was provided using the simple haptic sensation. 
	Under this condition, the same McKibben artificial muscles as in Condition 1 (proposed system) were attached to the arm instead of the torso for guidance. Fig.~\ref{fig:arm} illustrates the attachment of the artificial muscles to the arm. 
	\textcolor{black}{
		Table~\ref{tab:simple} shows the mapping between gaze guidance directions and the corresponding contracted artificial muscles illustrated in Fig.~\ref{fig:arm} for Condition 3.}
		The algorithm for determining the magnitude of the air pressure applied to the artificial muscles was the same as that for Condition 1 (proposed system), with the initial input air pressure determined by Equation~\eqref{eq:ap} and the air pressure adjusted by Equation~\eqref{eq:feedback}.
	In the air pressure adjustment used for Conditions 1 and 3, the adjustment gain $G$ in Equation \eqref{eq:feedback} was set to $0.05$.
	Further, under Condition 4, the synthesized speech was used for gaze guidance. 
	Under this condition, the subjects were guided by synthetic speech output from a neck speaker worn by the subject, which provided the target direction and the angle to the target angle. 
	For example, if the target direction is 36 deg to the right and 12 deg upward from the subject's gaze direction, the synthesized speech outputs ``36 to the right and 12 upward."
	Speech synthesis was output only once at the beginning of the experimental task.
	It should be noted that Microsoft Azure was used for speech synthesis in this study.
	\textcolor{black}{
		Synthetic speech guidance is expected to have the highest accuracy among these guidance methods, and a specific limitation of the proposed approach can be obtained by comparing the proposed system with synthetic speech guidance.
		No training was conducted to familiarize the participants with the conditions.
	}

	All 15 subjects were healthy individuals between the ages of 21 and 25, with their BMI ranging between 17 and 26.  
	Each subject performed the experimental task ten times under all conditions, and the order of conditions was determined randomly.
	
	\subsection{Experimental Results}
	\begin{figure}[!t]
		\centering
		\includegraphics[keepaspectratio=true,width=\linewidth]{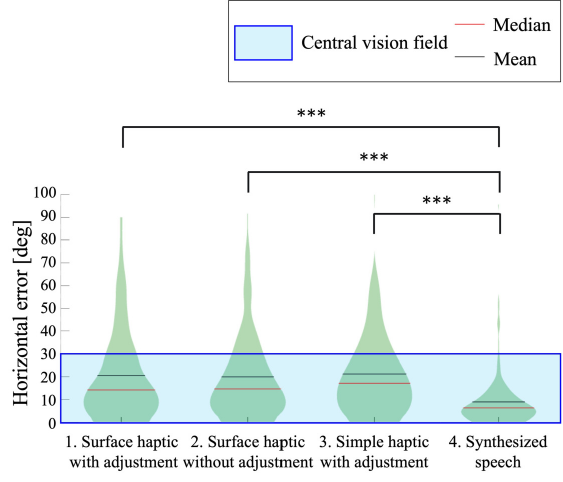}
		\vskip -3mm
		\caption{\textcolor{black}{
				Violin plot of the horizontal error. $***$ represents significant differences estimated by a multiple comparison (Friedman test with Dunn-Bonferroni correction \cite{dunn1964Multiple}: p $< 0.001$).
		}}
		\label{fig:violin_h}
		\vskip -3mm
	\end{figure}
	\begin{figure}[!t]
		\centering
		\includegraphics[keepaspectratio=true,width=\linewidth]{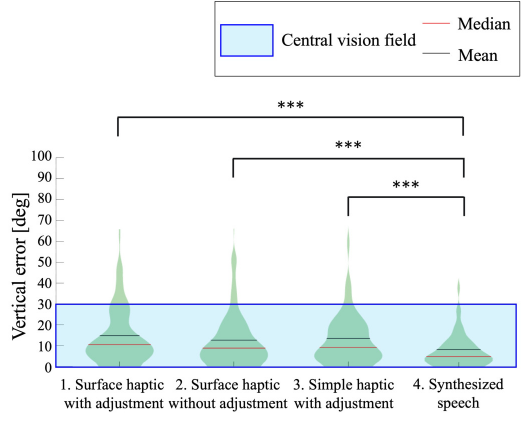}
		\vskip -3mm
		\caption{\textcolor{black}{
				Violin plot of the vertical error.  $***$ represents significant differences estimated by a multiple comparison (Friedman test with Dunn-Bonferroni correction \cite{dunn1964Multiple}: p $< 0.001$).
		}}
		\label{fig:violin_v}
		\vskip -3mm
	\end{figure}
	
	Figs.~\ref{fig:violin_h} and \ref{fig:violin_v} show the violin plots of the horizontal and vertical errors per condition.
	To provide gaze guidance, it is necessary to achieve an accuracy of 30 deg in radius from the center of the viewpoint, called the central vision field, which enables the accurate identification of the color and shape of an object~\cite{sampson2012Visual,bhise2011Ergonomics}.
	In all four conditions, the mean and median of the horizontal and vertical errors were less than 30 deg. Therefore, it was found that all four guidance methods, including the proposed method, were capable of guiding the gaze toward the central vision field. 
	
	Condition 4, which used synthesized speech, had significantly smaller horizontal and vertical errors than Conditions 1, 2, and 3, which used haptic guidance
	\textcolor{black}{ (Friedman test with Dunn-Bonferroni correction \cite{dunn1964Multiple}: p $< 0.001$).}
	Moreover, no significant differences were observed in the horizontal and vertical errors between Conditions 1, 2, and 3 with haptic guidance.
	
	\section{Experiment to Evaluate \textcolor{black}{ the Speed of Haptic Perception}}
	\subsection{Experimental Setup}
	The second experiment was conducted to evaluate the \textcolor{black}{speed of haptic perception}. 
	The subjects wore a mixed-reality device, wearable fabric actuator, and neck speaker, with six McKibben artificial muscles wrapped around each arm, as described in Section~\ref{sec:accuracy}.
	The subjects sat on a stool and placed a keyboard on their lap during the experiment.
	
\begin{figure}[!t]
		\centering
		\includegraphics[keepaspectratio=true,width=.9\linewidth]{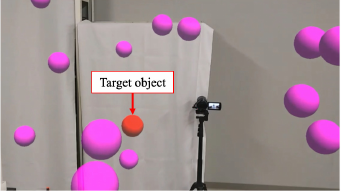}
		\caption{View from the subject's perspective during the experiment. The red sphere is the target object and the pink spheres are dummy objects. A target object was placed among 50 dummy objects.}
		\label{fig:view}
	\end{figure}

	Fig.~\ref{fig:view} shows a mixed-reality space view from a subject during the experiment.
	The subjects found as many target objects as possible for 40 s under four conditions described in Table~\ref{table:mr_condition}. 
	At the beginning of the experiment, the subjects faced forward and the target object with 50 dummy objects appeared randomly in the mixed-reality space. 	
	\textcolor{black}{
		Under conditions 1, 2, and 3, the air pressure was determined using Equation \eqref{eq:ap} according to the direction of the target object and the gaze angle of the subject. Under conditions 1 and 3, the air pressure was adjusted using Equation \eqref{eq:feedback} until the target object was located. Under Condition 2, the same air pressure was maintained until the target object was located. Under Condition 4, speech synthesis was output only once for each target object.
	}
	The subjects searched for the target object and pressed the keyboard upon finding it. 
	When the subject presses the keyboard, the error angle between the subject's gaze direction and the target direction is measured. 
	If this error angle is within 5 deg, the dummy object disappears and the target object returns to the subject's front. 

\begin{figure}[!t]
		\centering
		\includegraphics[keepaspectratio=true,width=\linewidth]{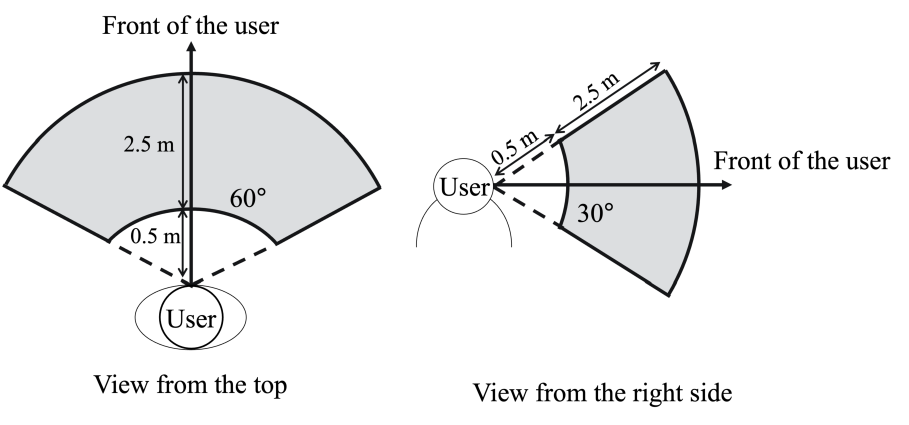}
		\caption{\textcolor{black}{Display range of virtual objects. Target and dummy objects appear within the gray regions.}}
		\label{fig:top}
	\end{figure}

		The subject then had to go back gazing at the starting position, the subject's front. 
		When the subject looks at the starting position, the target object randomly moves again and dummy objects appear.
	\textcolor{black}{
	Fig.~\ref{fig:top} shows the display range of the target and dummy objects from the top and right sides of the subject. }
	The virtual objects were displayed within a horizontal angle of 60 deg to the left or right, a vertical angle of 30 deg above or below, and a distance of 0.5 to 3.0 m from the subject, denoted by gray areas in Fig.~\ref{fig:top}.
	Both the target object and dummy objects were spheres with a diameter of 0.1 m; the color code of the target and dummy objects was \#FF0000 and \#FF00FF, respectively. 
	\textcolor{black}{
		In most of the experimental tasks, the target object was not in the visual field of the subject before guidance. Each guidance method assists the subject by moving the subject's central visual field and placing the target object within it.
		Changing the color of the target object allows participants to easily identify that the target object has entered their central visual field. The experiment was designed based on previous studies \cite{grogorick2017Subtle,nakamura2022Evaluation}.
	}
	The \textcolor{black}{speed of haptic perception} was evaluated by measuring the number of identified target objects.
	
	All 15 subjects were healthy individuals between the ages of 21 and 25, with their BMI ranging between 17 and 26, as previously described in Section~\ref{sec:accuracy}. 
	Each subject performed the experimental task six times under all conditions, and the order of conditions was randomly determined.
	
	\subsection{Experimental Results}
	\begin{figure}[!t]
		\centering
		\includegraphics[keepaspectratio=true,width=\linewidth]{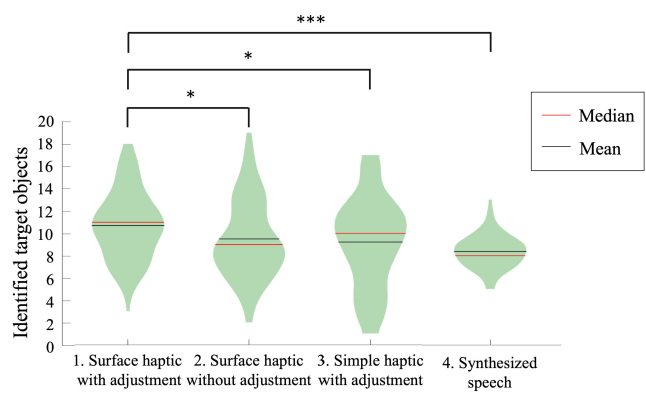}
		\caption{\textcolor{black}{
				Violin plot of identified target objects. $***$ represents significant differences by a multiple comparison (Friedman test with Dunn-Bonferroni correction \cite{dunn1964Multiple}: p $< 0.001$).
				$*$ represent significant differences by a multiple comparison (Friedman test with Dunn-Bonferroni correction \cite{dunn1964Multiple}: p $< 0.05$).
		}}
		\label{fig:violin_speed}
	\end{figure}
	\begin{table}[!t]
		\centering
		\caption{Mean Value of Identified Target Objects per Subject.}	
		\label{tab:Mean_speed}
		\begin{tabular}{c|cccc}
			Guidance method& \begin{tabular}[c]{@{}c@{}}Surface\\ haptic\end{tabular} & \begin{tabular}[c]{@{}c@{}}Surface\\ haptic\end{tabular} & \begin{tabular}[c]{@{}c@{}}Simple\\ haptic\end{tabular} & \begin{tabular}[c]{@{}c@{}}Synthesized\\ speech\end{tabular}    \\\hline
			Air pressure  adjustment& With & Without   & With &Without \\\hline
			A & 7.7 & 8.7 & 7.8   & 6.5  \\
			B & 11.8 & 11.8  & 8.3   &7.8     \\
			C & 10.8 & 9.3& 10.0 &  8.7     \\
			D & 11.3  & 12.0 & 11.2  &  8.8     \\
			E & 15.3  & 11.5 & 14.7 & 9.5     \\
			F & 10.7  & 10.0   & 10.0 & 8.7     \\
			G & 7.2   & 6.2 & 5.2 &  8.0     \\
			H & 10.7  & 10.0 & 4.3   &  6.5     \\
			I  & 10.5& 8.0 & 8.5 & 8.3     \\
			J  & 8.0 & 7.7  & 5.8  & 7.5  \\
			K  & 10.3  & 6.7  & 8.7   & 8.2     \\
			L  & 13.5  & 9.0 & 13.3  & 10.5    \\
			M & 14.5 & 16.2  & 13.2  & 10.7    \\
			N   & 8.7& 7.8 & 9.2   &  8.7     \\
			O    & 9.7 & 7.7 & 8.2& 7.3 
		\end{tabular}
	\end{table}
	Fig.~\ref{fig:violin_speed} illustrates the violin plot of the number of identified target objects per condition.
		The standard deviations of the identified target objects under conditions 1, 2, 3, and 4 were 3.0, 3.3, 4.0, and 1.4, respectively.
	Table~\ref{tab:Mean_speed} presents the mean of the number of identified target objects per subject, respectively. In this table, A through O represent each subject.
	
	The proposed method (Condition 1) demonstrated the largest mean among all four conditions.
	Under Condition 1 (the proposed method), ten subjects had the largest mean. 
	Under Condition 2, four subjects had the largest mean. 
	Under Condition 3, one subjects had the largest mean. 
	Under Condition 4 one subjects had the largest mean. 
	
	Multiple comparison revealed that Condition 1 (the proposed method) had significantly higher identified target objects than Conditions 2, 3, and 4, 
	\textcolor{black}{ (Friedman test with Dunn-Bonferroni correction \cite{dunn1964Multiple}: p $< 0.05$).}
	The results indicate that the proposed method can guide the gaze direction at the highest speed among all four conditions. 
	
	\section{Demonstration of Proposed System in Immersive Micromanipulation System}
	\begin{figure}[!t]
		\centering
		\includegraphics[keepaspectratio=true,width=\linewidth]{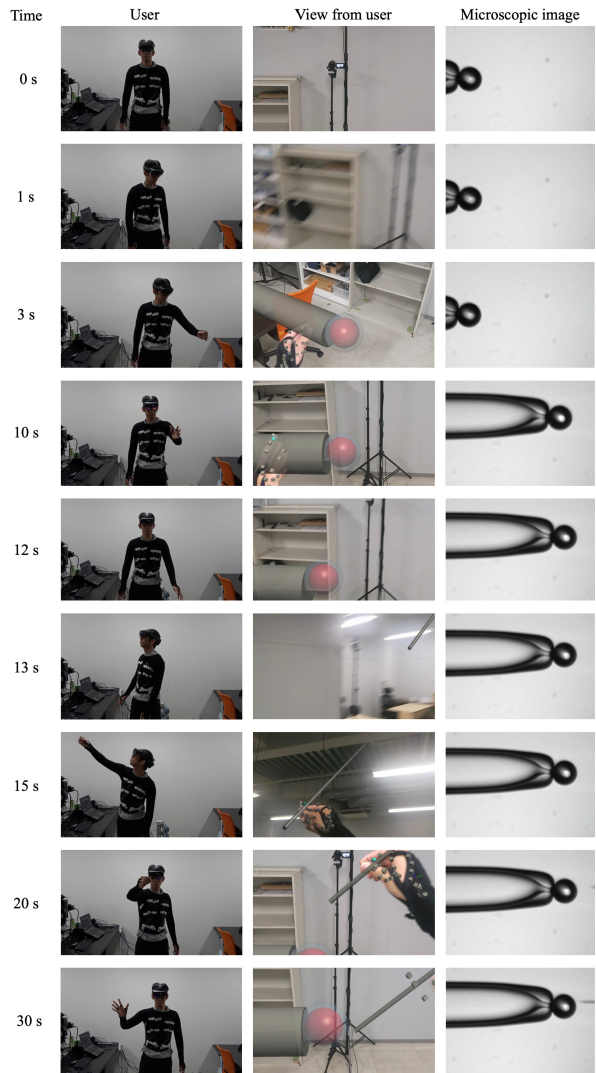}
		\caption{Demonstration of the directional sense presentation in an immersive micromanipulation system. The user wears a wearable fabric actuator and a mixed-reality device. Mixed reality device displays virtual objects to the user. The virtual objects are a holding pipette, an injection pipette, and a microbead, and the user performs micromanipulation by operating the virtual pipettes. Based on the position of the operated virtual pipettes, the actual pipettes in microscopy are operated.}
		\label{fig:demo_fabric}
	\end{figure}
	A demonstration was conducted to confirm the application of directional sense presentation using a wearable fabric actuator to an actual immersive manipulation system.
	In this demonstration, we implemented directional sense presentation using a wearable fabric actuator in a mixed-reality-based immersive micromanipulation system proposed by our research group~\cite{yokoe2022immersive,yokoe2022MixedRealityBased} and confirmed whether the system can smoothly perform micromanipulation. 
	This immersive micromanipulation system enabled the operator to perform micro-manipulation by moving virtual pipettes. 
		By properly mapping the micropipettes and micro-objects onto a macro 3D immersive space, this system allows the user to perform micromanipulation with hand movements 1000 times larger than the actual micromanipulation task. Therefore, users can perform micromanipulation as if the tasks were performed in the macro world. However, the user may lose sight of the micropipette and micro-objects to be manipulated because the user is immersed in an immersive space. The proposed guidance system is then used to guide the user to the location of micropipettes or micro-object.
	
	\textcolor{black}{Fig.~\ref{fig:demo_fabric} shows a demonstration of the proposed system in micromanipulation tasks.}
	At the beginning of the demonstration, the operator did not see the virtual holding pipette or microbeads to be manipulated. Therefore, the wearable fabric actuator guided the operator to find the virtual holding pipette (Time = 0--3 s in Fig.~\ref{fig:demo_fabric}).
	The operator smoothly found the virtual holding pipette and microbeads and manipulated them using his hand (Time = 3--12 s in Fig.~\ref{fig:demo_fabric}).
	The operator then wanted to manipulate the injection pipette but could not see the virtual injection pipette.
	Therefore, the wearable fabric actuator provided the operator with the direction of the virtual injection pipette (Time = 12--15 s in Fig.~\ref{fig:demo_fabric}).
	Then, the operator smoothly found and manipulated the virtual injection pipette (Time = 15-30 s in Fig.~\ref{fig:demo_fabric}).
	
	This demonstration confirmed that the proposed directional sense presentation system can be implemented in an actual immersive manipulation system.
	
	\section{Discussion}
	In the proposed derivation method of an equation between air pressure and gaze angle, air pressure is applied to the wearable fabric actuator in the order of smallest to largest.
	However, the hysteresis of surface haptic sensation produced by the wearable fabric actuator may cause users to move differently from the directional presentation that is actually performed in an immersive space.
	Therefore, we implemented an air adjustment algorithm that gradually changes, as in the derivation process, the air pressure when users make an incorrect move to improve the directional guidance performance.
	The experiment for evaluating the speed of directional sense presentation confirmed that the adjustment algorithm significantly increased the speed of directional sense presentation using surface haptic sensation.
	It is known that the human perception of haptic sensation is relative~\cite{weber2018Weber}; therefore, changes in the intensity of surface haptic sensation by the adjustment algorithm may have helped users to perceive the intensity of surface haptic sensation more easily, thereby increasing the directional guidance speed of surface haptic sensation.
	
	The proposed method also had a significantly higher speed of directional guidance than simple haptic sensation, however, there is no significant difference in the speed between surface haptic sensation without air pressure adjustment and simple haptic sensation with the adjustment. 
	This result suggests that the proposed air pressure adjustment method only affected the surface haptic sensation of the wearable fabric actuator.
	Since the surface haptic sensation generated by the wearable fabric actuator followed body movements, the wearer experienced greater resistance than that in a simple haptic sensation when performing a movement different from the guided movement.
	The proposed air pressure adjustment changes the intensity of surface haptic sensation when the wearer makes a movement different from the guided movement, and the wearer experiences greater resistance earlier than without the air pressure adjustment method.
	It is believed that the large resistance increased the speed of directional sense presentation by making the wearer quickly recognize that the movement was different from the guided movement.
	The intensity of haptic sensation can be perceived using a simple haptic sensation; however, it is difficult to perceive resistance when the user makes a movement that is different from the guided movement; thus, air pressure adjustment may be less effective for simple haptic sensation. 
	This sense of resistance makes it easier for subjects to recognize when they have made a large error in their movements, therefore the standard deviation for the conditions using surface haptic sensation were lower than that for the simple haptic sensation condition in the experiment evaluating the speed of directional sense presentation.
	
	The proposed system was confirmed to have outstanding speed in presenting a directional sense. 
	However, when the proposed system was used in the experiment to evaluate the speed of directional sense presentation, five of the 15 subjects did not have the highest mean of the identified target objects. 
	\textcolor{black}{
		A few subjects did not obtain a good surface haptic sensation because the proposed wearable fabric actuator did not match their body size.
		In this study, the wearable fabric actuator was available in only one size; therefore, it may have been too large for some of the subjects. 
		Although the calibration steps proposed in this study can address differences in the subject's sensitivity to haptic sensations, they cannot address differences in the subject's body size. In future, we plan to develop fabric actuators of various sizes to address the body size of the wearer.
	}
	Additionally, in the subject experiments, the adjustment gain value was kept constants for all subjects. However, the adjustment gain that maximized the adjustment effect differed among the subjects, and the performance of the proposed system may be further improved by adjusting the adjustment gain of each user. 
	Moreover, for subject safety, the magnitude of air pressure applied to the artificial muscles was limited to 375 kPa in the subject experiment. The air pressure did not change when the air pressure adjustment calculation results were higher than 375 kPa. 
	Therefore, subjects insensitive to surface haptic sensation would not improve the speed of directional sense presentation because the air pressure reached the upper limit and did not change. 
	
	In the evaluation of directional presentation accuracy, the accuracy of directional guidance using haptic sensations were significantly lower than that using synthesized speech. 
	Therefore, the synthesized speech was less misleading to users and facilitated the understanding of the direction of guidance in comparison to haptic sensations.
	This result may be because humans communicate using speech daily and are used to transforming speech into directional information; however, they are not familiar with transforming haptic sensation into directional information. For this reason, in the experiment evaluating the speed of directional sense presentation, the standard deviation in conditions using synthesized speech was lower than that in conditions using haptic sensation.
	If humans become familiar with transforming haptic sensation into directional information, the accuracy of haptic-based directional presentation may improve. 
	
	The \textcolor{black}{speed of haptic perception} using the proposed method was significantly higher than using synthesized speech. 
	Furthermore, in the evaluation of directional presentation speed, the mean values of identified target objects using synthesized speech was lower than the values using haptic sensations. 
	The results could be explained by the fact that synthesized speech requires more time to receive information and transform the information into direction than haptic sensation.
	
	In the evaluation of directional presentation accuracy, there were no combinations with significant differences between the conditions using haptic sensations. 
	This result suggests that the proposed air pressure adjustment method does not have a significant effect on the accuracy of haptic directional sense presentation. 
	In addition, the location of the haptic sensation may have no significant effect on accuracy if the magnitude of the haptic sensation remains the same.
	
	\textcolor{black}{
		This study proposed an air pressure adjustment algorithm for the directional guidance of static targets; therefore, the proposed system cannot be directly applied to the directional guidance of moving targets. However, the proposed system can be applied to moving targets by developing an algorithm that adjusts the air pressure according to the movement of the target.
	}
	
	\section{Conclusion}
	In this study, we proposed a directional sense presentation system in an immersive space using a wearable fabric actuator. 
	The proposed system guides the user's direction by providing surface haptic sensation to the user according to the user's direction.
	Furthermore, a novel air pressure adjustment method for surface haptic sensation based on human motion was implemented in the proposed system.
	We conducted experiments to verify the effectiveness of the proposed system, which confirmed that the proposed system can provide directional guidance significantly faster than synthesized speech guidance and simple haptic sensation. 
	The experimental results also indicate that the proposed air pressure adjustment method for surface haptic sensation improved the \textcolor{black}{speed of haptic perception}.
	Finally, we demonstrate the proposed system using a mixed-reality-based micromanipulation system to confirm that it can be implemented in an actual immersive manipulation system. 
	
	The directional presentation by surface haptic sensation could be adapted to other immersive systems and could dramatically increase the speed of work in immersive spaces without disturbing the user's movement in the immersive space.
	Future research objectives include verifying the correlation between users' familiarity with the wearable fabric actuator and the efficiency of directional guidance, as well as developing a surface haptic sensation presentation method for motion guidance in other parts of the body.


	
	%
	\bibliography{tohfull}
	\bibliographystyle{IEEEtran}
	
	%
	%
	
	\begin{IEEEbiography}[{\includegraphics[width=1in,height=1.25in,clip,keepaspectratio]{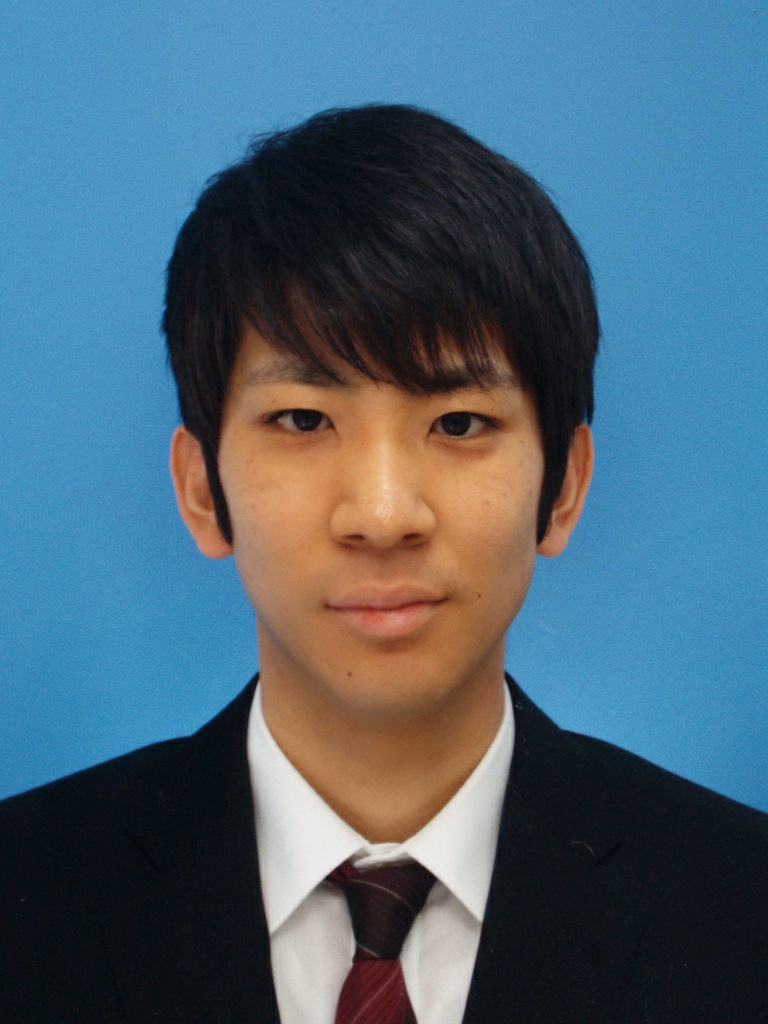}}]{Kenta Yokoe}
		(Graduate Student Member, IEEE) received the B.E. degree in mechanical engineering and the M.E. degree in micro–nano mechanical science and engineering from Nagoya University, Nagoya, Japan, in 2021 and 2023, respectively, where he is currently pursuing the Ph.D. degree. in the Department of Micro-Nano Mechanical Science and Engineering at Nagoya Universi His research interests include human–machine interaction and haptic technology.
	\end{IEEEbiography}

	\begin{IEEEbiography}[{\includegraphics[width=1in,height=1.25in,clip,keepaspectratio]{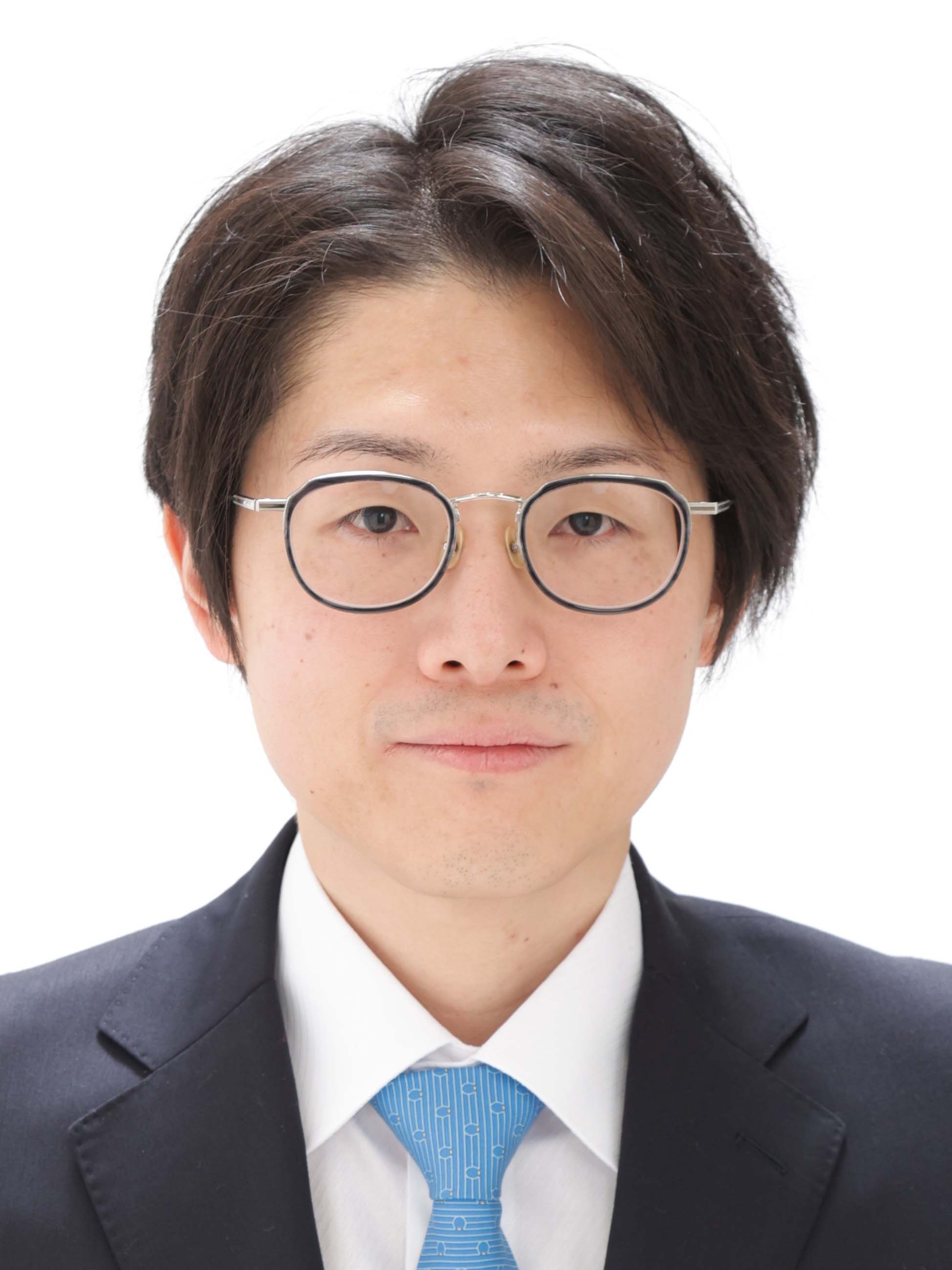}}]{Tadayoshi Aoyama}
		(Member, IEEE) received his B.E. degree in Mechanical Engineering, M.E. degree in Mechanical Science and Engineering, and Ph.D. degree in Micro-Nano Systems Engineering from Nagoya University, Japan, in 2007, 2009, and 2012, respectively. He served as an Assistant Professor at Hiroshima University (2012–2017), and at Nagoya University (2017–2019). He then became an Associate Professor at Nagoya University (2019–2024). From 2018 to 2022, he was a PRESTO Researcher at JST. He is currently a Professor in the Department of Mechanical Systems Engineering at Nagoya University. His research focuses on macro-micro interaction, VR/AR and human interfaces, AI-based assistive technology, micromanipulation, and medical robotics.
	\end{IEEEbiography}

	\begin{IEEEbiography}[{\includegraphics[width=1in,height=1.25in,clip,keepaspectratio]{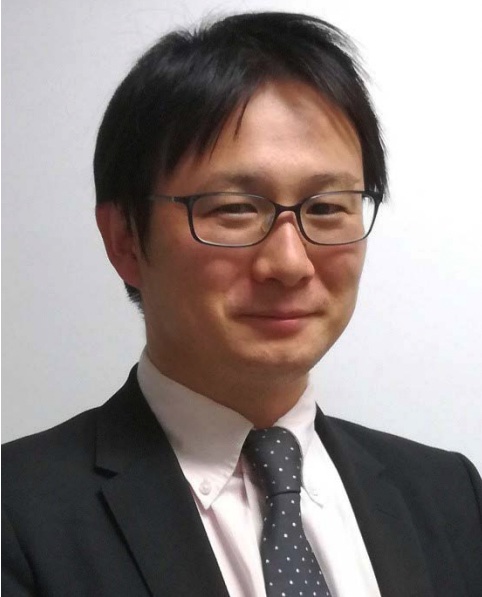}}]{Yuki Funabora}
		(Member, IEEE) received B.E., M.E., and Ph.D. degrees in electrical engineering and computer science from Nagoya University, Nagoya, Japan, in 2007, 2009, and 2012, respectively. In 2012, he was a Postdoctoral Researcher at the RIKEN Advanced Science Institute. He was an Assistant Professor at Nagoya University from 2013 to 2020. He was also a PRESTO Researcher at JST from 2018 to 2022. He is currently an Associate Professor in the Department of Information and Communication Engineering of at Nagoya University and a FOREST researcher at JST since 2020 and 2022, respectively. His research interests include human-cooperative robots, soft robotics, intelligent control, soft computing, and system design.
	\end{IEEEbiography}

	\begin{IEEEbiography}[{\includegraphics[width=1in,height=1.25in,clip,keepaspectratio]{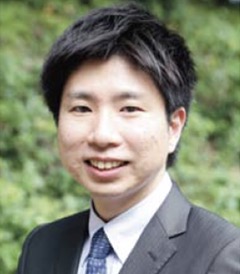}}]{Masaru Takeuchi}
		(Member, IEEE) received M.E. and Ph.D. degrees in micro–nano systems engineering from Nagoya University, Nagoya, Japan, in 2010 and 2013, respectively. He studied at the University of California, Los Angeles, CA, USA, for a year starting in June 2011. From April 2013 to June 2014, he worked as a Postdoctoral Research Fellow at Nagoya University, Japan. In July 2014, he became an Assistant Professor at the Department of Micro-Nano Systems Engineering at Nagoya University. He was Designated Assistant Professor with the Institute for Advanced Research at Nagoya University from April 2016 to April 2020. He was also a visiting researcher in Plastic Surgery at the University of Michigan, MI, USA, from March 2016 to June 2017. He has been an Assistant Professor with the Department of Micro–Nano Mechanical Science and Engineering at Nagoya University and a FOREST researcher at JST since May 2020 and April 2021, respectively. His main research interests include implantable medical devices, cyborg technology, and micro–mechatronics.
	\end{IEEEbiography}

	\begin{IEEEbiography}[{\includegraphics[width=1in,height=1.25in,clip,keepaspectratio]{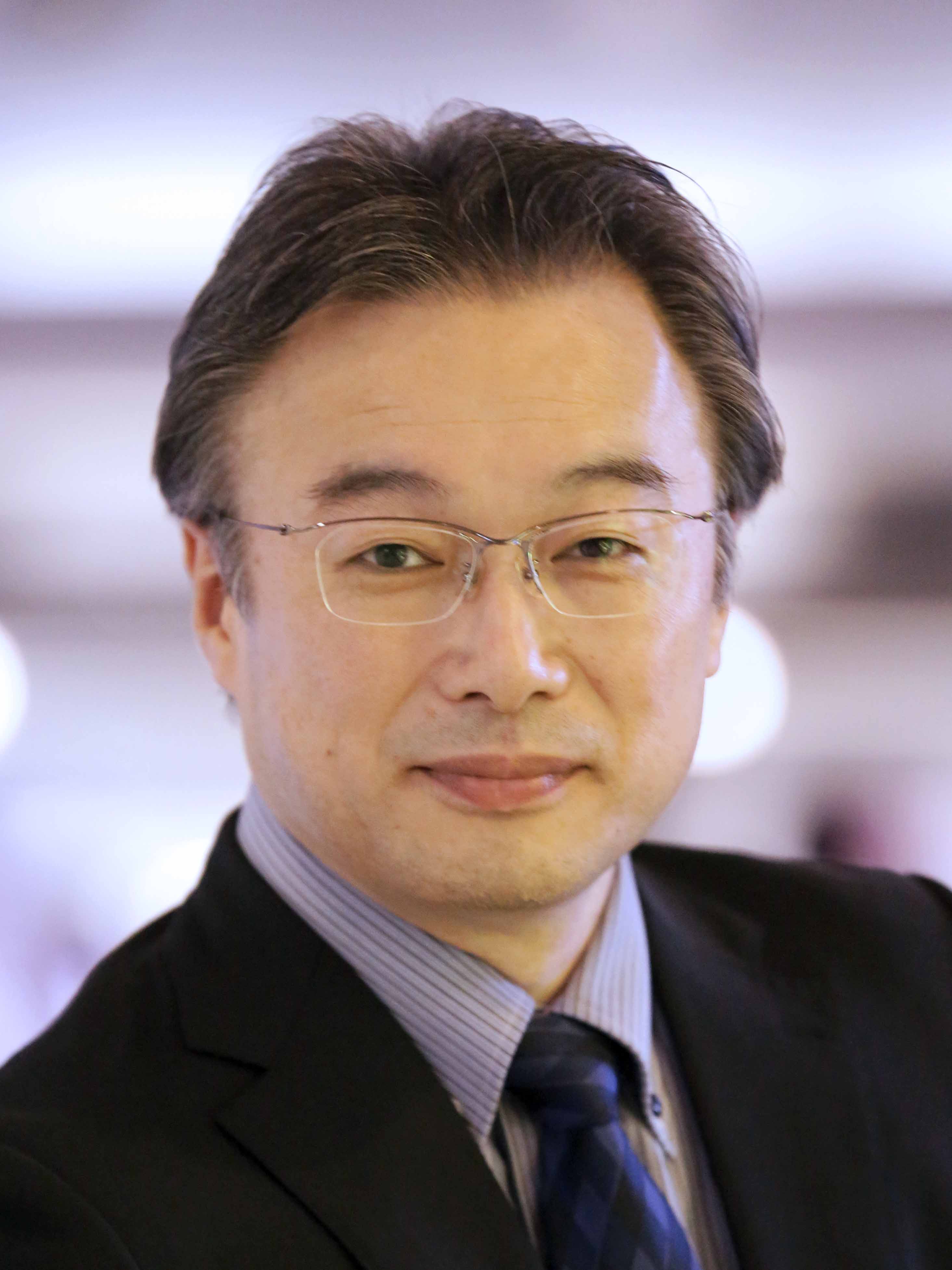}}]{Yasuhisa Hasegawa}
		(Member, IEEE) received B.E., M.E., and Ph.D. degrees in Robotics from Nagoya University, Nagoya, Japan, in 1994, 1996, and 2001, respectively. From 1996 to 1998, he worked at Mitsubishi Heavy Industries Ltd., Japan. He joined the Nagoya University in 1998. He then moved to Gifu University in 2003. From 2004 to 2014, he attended the University of Tsukuba. Since 2014, he has been with Nagoya University, where he is currently a professor in the Department of Micro–Nano Mechanical Science and Engineering. His primary research interests include motion-assistive systems, teleoperation for manipulation, and surgical support robots.
	\end{IEEEbiography}


	\vfill
	
\end{document}